\documentclass[12pt]{article}
\usepackage[dvips]{graphicx}
\begin{document}
\topmargin -10mm
\title{Quasinormal frequencies of Schwarzschild black holes in 
anti-de Sitter spacetimes: A complete study on the 
overtone asymptotic behavior}

\author{Vitor Cardoso\\
Centro Multidisciplinar de Astrof\'{\i}sica - CENTRA,\\
Departamento de F\'{\i}sica, Instituto Superior T\'ecnico,\\
Av. Rovisco Pais 1, 1049-001 Lisboa, Portugal\\
vcardoso@fisica.ist.utl.pt\\ \\
Roman Konoplya \\
Department of Physics, Dniepropetrovsk National University\\
St. Naukova 13, Dniepropetrovsk  49050, Ukraine\\
konoplya$_{-}$roma@yahoo.com\\ \\
Jos\'e P. S. Lemos\\
Department of Physics, Columbia University, \\
New York, NY 10027, \&\\
Centro Multidisciplinar de Astrof\'{\i}sica - CENTRA,\\
Departamento de F\'{\i}sica, Instituto Superior T\'ecnico,\\
Av. Rovisco Pais 1, 1049-001 Lisboa, Portugal, \\
lemos@physics.columbia.edu}
\date{}
\maketitle
\thispagestyle{empty}
\newpage
\centerline{}
\vskip 4cm
Abstract: We present a thorough analysis for the quasinormal (QN) behavior,
associated with the decay of scalar,
electromagnetic and gravitational perturbations, of 
Schwarzschild black holes in anti-de Sitter (AdS) spacetimes. 
As it is known the AdS QN spectrum crucially depends on the relative
size of the black hole to the AdS radius.  There are three
different types of behavior depending on whether the
black hole is large, intermediate, or small.  The results of previous
works, concerning lower overtones for large black holes, are completed
here by obtaining higher overtones for all the three black hole
regimes.  There are two major conclusions that one can draw from this
work: First, asymptotically for high overtones, all the modes are
evenly spaced, and this holds for all three types of regime, large,
intermediate and small black holes, independently of $l$, where $l$ is
the quantum number characterizing the angular distribution; Second,
the spacing between modes is apparently universal, in that it does not
depend on the field, i.e., scalar, electromagnetic and
gravitational QN modes all have the same spacing for high
overtones.  We are also able to prove why scalar and gravitational
perturbations are isospectral, asymptotically for high overtones, by
introducing appropriate superpartner potentials.

\newpage

\section{Introduction}

Any physical system has its modes of vibrations.  For non-dissipative
systems these modes, forming a complete set, are called normal. Each
mode having a given real frequency of oscillation and being
independent of any other. The system once disturbed continues to
vibrate in one or several of the normal modes. On the other hand, a
black hole, as any other gravitational system, is a dissipative system
since it emits gravitational radiation. One has to consider,
instead, quasinormal modes (QNMs) for which the frequencies are no
longer purely real, showing that the system is loosing energy.  QNMs
are in general not complete and though insufficient to fully describe
the dynamics, contain a great amount of information. For instance,
they dominate the signal during the intermediate stages of the
perturbation. Indeed, calculations ranging from the formation of a
black hole in gravitational collapse to the collision of two black
holes provide clear evidence that no matter how one perturbs a black
hole, its response will be dominated by the QNMs. Through the QNMs,
one can also probe the black hole mass, electric charge and angular
momentum by inspection of their characteristic waveform, as well as
test the stability of the event horizon against small
perturbations. Moreover, the interest in QNMs has now broaden, they
might be related to fundamental physics. In the context of black holes
in asymptotically flat spacetimes the importance of QNMs has long ago
been recognized.  Due to a crescent increase of interest for black
holes in asymptotically de Sitter (dS) and asymptotically anti-de
Sitter (AdS) spacetimes the study of QNMs has now spread into these.

In asymptotically flat spacetimes the idea of QNMs started with the
work of Regge and Wheeler \cite{reggeW} where the stability of a black
hole was tested, and were actually first numerically computed by
Chandrasekhar and Detweiler several years later \cite{Chandra1}. It
continues to be a very active field due to the eminent possibility of
detecting gravitational waves from astrophysical sources.  There are
two standard reviews in the field \cite{kokkotasnollert1}.  QNMs in
asymptotically flat spacetimes have recently acquired a further
importance since it has been proposed that the Barbero-Immirzi
parameter, a factor introduced by hand in order that Loop Quantum
Gravity reproduces correctly the black hole entropy, is equal to the
real part of the quasinormal (QN) frequencies with a large imaginary part
\cite{hoddreyerkunstattercorichimotldreyer2} (see \cite{baez} for a
short review).  For further developments in calculating QN frequencies
in Kerr spacetimes and in asymptotically flat
black holes spacetimes in d-dimensions see \cite{kokoko}.
Some other recent calculations of QN frequencies
in asymptotically flat black hole spacetimes can be found in 
\cite{konoplya2}.

In asymptotically dS spacetimes the calculation of QN frequencies was
first done by Moss and collaborators \cite{mm1}
in which the stability of the Cauchy horizon of a charge black hole was
analyzed. In Cardoso and Lemos 
\cite{cardosolemos1} an analytical method was devised to
study the case in which the black hole and the cosmological horizons
are very close to each other. 
This analytical method has recently been extended to higher dimensional
Schwarzschild dS black holes \cite{molina} and also to higher order
in the difference between the cosmological and horizon radius \cite{brink}.
A different analytical approach has also been developed in \cite{suneeta}.

In asymptotically AdS spacetimes, which is the background spacetime
for our paper, there has been a great amount of work because the
AdS/CFT correspondence conjecture \cite{maldacena} makes the
investigation of QNMs important.  According to it, the black hole
corresponds to a thermal state in the conformal field theory (CFT),
and the decay of the test field in the black hole spacetime,
corresponds to the decay of the perturbed state in the CFT.  The
dynamical timescale for the return to thermal equilibrium can be done
in AdS spacetime, and then translated onto the CFT, using the AdS/CFT
correspondence.  Many authors have now delved into these calculations
in several different black hole settings in several different
dimensions (see \cite{qn} for a sample). In this paper we are
interested in the 4-dimensional Schwarzschild-AdS spacetime.  The
lowest lying modes (i.e., the less damped ones parametrized by the 
overtone number $n=0$) for this spacetime were found by Horowitz and
Hubeny \cite{horowitz}, and completed by Konoplya \cite{konoplya} for
the scalar case, and by Cardoso and Lemos \cite{cardosolemos2} for the
electromagnetic and gravitational case.  Recently, Berti and Kokkotas
\cite{berti} have confirmed all these results and extended them to
Reissner-Nordstr\"om-AdS black holes. Here, we shall take a step
further in carrying on this program by computing numerically, through
an extensive search, the high overtone QN frequencies for scalar,
electromagnetic and gravitational perturbations in the
Schwarzschild-AdS black hole. We shall do an extensive search for the
high overtone QN frequencies, ($n\geq1$). We find that the modes are 
evenly spaced
for frequencies with a large imaginary part.  Moreover the scalar,
electromagnetic and gravitational perturbations all possess,
asymptotically for high overtones $n$, QN frequencies with the same
spacing, and this spacing is $l$-independent.  While we can
numerically prove this with great accuracy for large black holes, it
remains just a conjecture for small and intermediate black holes.
We shall also see that the QN frequencies of the toroidal black hole
with non-trivial topology \cite{lemos} are identical to the QN
frequencies of a large Schwarzschild-AdS black hole
\cite{cardosolemos3}. 

\section{Equations and Numerical Method}
We shall deal with the free evolution of massless classical 
fields in the
background of a Schwarzschild-AdS spacetime with metric
given by
\begin{equation}
ds^{2}= f(r) dt^{2}- \frac{dr^{2}}{f(r)}-
r^{2}(d\theta^{2}+\sin^2\theta d\phi^{2})\,,
\label{lineelement}
\end{equation}
where, $f(r)=\frac{ r^{2}}{R^2}+1-\frac{2M}{r}\,$, $R$ is the AdS
radius, and $M$ the black hole mass (Newton's constant $G_{\rm N}$
and the velocity of light are set to one).  The evolution of scalar,
electromagnetic, and gravitational fields can be followed through the
Klein-Gordon, Maxwell and Einstein equations, 
respectively. If one assumes that the fields are a small perturbation
in the background given by equation 
(\ref{lineelement}), then all the covariant
derivatives can be taken with respect to the metric 
(\ref{lineelement}). It is then 
possible to show that the evolution equations are all of the same
type, i.e., a second order radial differential equation (for more
details we refer the reader to \cite{horowitz} for
the scalar case and to \cite{cardosolemos2} for the
electromagnetic and gravitational case). The wave equation is  
\begin{equation}
\frac{\partial^{2} \Psi(r)}{\partial r_*^{2}} +
\left\lbrack\omega^2-V(r)\right\rbrack\Psi(r)=0 \,,
\label{waveeq}
\end{equation}
where the tortoise coordinate $r_*$ is defined as
\begin{equation}
\frac{\partial r}{\partial r_*}= f(r)\,,
\end{equation}
and the potential $V$ appearing in (\ref{waveeq}) depends on the
specific field under consideration.
Explicitly, for scalar perturbations,
\begin{equation}
V_{\rm s}=f(r)\left\lbrack\frac{l(l+1)}{r^2}+
\frac{2M}{r^3}+\frac{2}{R^2}\right\rbrack \,,
\label{Vscalar}
\end{equation}
while for electromagnetic perturbations,
\begin{equation}
V_{\rm em}=f(r)\left\lbrack\frac{l(l+1)}{r^2}\right\rbrack \,.
\label{Velectromagnetic}
\end{equation}
The gravitational perturbations decompose into two sets 
\cite{cardosolemos2},
the odd and the even parity one.
For odd perturbations the potential $V(r)$ in (\ref{waveeq}) is
\begin{equation}
V_{\rm odd}=  f(r)
\left\lbrack\frac{l(l+1)}{r^2}-\frac{6M}{r^3}\right\rbrack\,.
\label{vodd}
\end{equation}
while for even perturbations, we have
\begin{equation}
V_{\rm even}= \frac{2f(r)}{r^3}
\frac{9M^3+3\alpha^2Mr^2+\alpha^2\left(1+\alpha\right)r^3+3M^2
\left(3\alpha r+3\frac{r^3}{R^2}\right)}
{\left(3M+\alpha r\right)^2} \,,
\label{veven}
\end{equation}
where $\alpha=\frac{1}{2}\left\lbrack l(l+1)-2\right\rbrack$.  In all
cases, we denote by $l$ the angular quantum number, that gives the
multipolarity of the field.  We can of course rescale $r$,
$r\rightarrow\frac{r}{R}$. If we do this, the wave equation 
takes again the form (\ref{waveeq}) with rescaled constants i.e., $r_+
\rightarrow \frac{r_+}{R}$, $\omega \rightarrow \omega R $, where
$r_+$ is the horizon radius.  So, we can take $R=1$ and measure
everything in terms of $R$, the AdS radius. 
Eq. (\ref{waveeq}) should be solved under
appropriate boundary conditions, i.e., incoming waves near the horizon,
\begin{equation}
\Psi \sim e^{-i\omega r_*}\,,\, r \rightarrow r_+\,,
\label{b1}
\end{equation}
and no waves at infinity, 
\begin{equation}
\Psi=0\,,\,r \rightarrow \infty\,.
\label{b2}
\end{equation}
We note
that there are other reasonable boundary conditions at infinity, in
particular for the gravitational perturbations. For instance, 
one can define
Robin boundary conditions in such a way as to preserve certain
dualities between the odd and the even gravitational perturbations. 
However, it was verified numerically by Moss and Norman
\cite{mm1} that Dirichlet or Robin boundary conditions yield
approximately the same result, so we shall keep $\Psi=0\,,\,r
\rightarrow \infty$. Moreover, Cardoso and Lemos \cite{cardosolemos2}
proved that for high overtone QN frequencies the duality
is preserved, so in this regime the distinction is irrelevant.
Thus, to compute the
QN frequencies $\omega$ such that the boundary conditions
(\ref{b1}) and (\ref{b2}) are preserved, we follow the 
Horowitz-Hubeny approach \cite{horowitz}. Within this approach we need to
expand the solution to the wave equation around
$x_{+}=\frac{1}{r_{+}}$ ($x=1/r$), 
\begin{equation}\label{my1}
\Psi (x)=\sum_{k=0}^{\infty} a_{k}(\omega)(x-x_{+})^{k}\,,
\end{equation}
and to find the roots of the equation $\Psi(x=0)=0$. First, one
should substitute  (\ref{my1}) into the wave equation (\ref{waveeq})
in order to obtain a recursion relation for $a_{k}$ \cite{horowitz}.
Then, one has to truncate the
sum (\ref{my1}) at some large $k=N$ and check that for greater $k$ the
roots converge to some true root which is the sought QN frequency.
The higher the overtone number, and the smaller the
black hole size, the larger the number $N$ at which the
roots of the equation $\Psi(x=0)=0$ converge. Yet, since in the
series (\ref{my1}) each next term depends on all the preceding terms
through the recursion relations, when $N$ is too large, the tiny
numerical errors in the first terms, start growing as $N \sim
10^2-10^3$ or greater. As a result the roots suffer a sharp change for a
small change on any of the input parameters, displaying a ``noisy''
dependence. To avoid this we have to increase the precision of all the
input data and the recursion relation we are dealing with from the
standard 20-digit precision up to a  precision such that further
increasing of it will not influence the result for the QN
frequency. For small black holes the roots start converging 
at very large $N$ only,
for instance, when $r_{+}=1/20$ we can truncate the series at $N
\sim 3\cdot10^4$, but not before. 
Since for finding roots of (\ref{b2}) we have to
resort to trial and error method, the above procedure consumes much
time, but nevertheless allows to compute QNMs of small
AdS black holes \cite{konoplya}, and to find the higher overtones we are
seeking.

\section{Numerical Results}
In this section we will present the numerical results obtained using
the numerical procedure just outlined in the previous section.  The
results will be organized into three subsections: scalar,
electromagnetic and gravitational perturbations. 
For each field, we shall also
divide the results into three different regimes:  large,
intermediate and small black holes, since  the results depend
crucially on the regime one is dealing with.  Here a large black
hole stands for a black hole with $r_+ \gg 1$, an intermediate black
hole is one with $r_+ \sim 1$, and a small black hole has a horizon
radius $r_+ \ll 1$.  We shall then try to unify these results.  For
each horizon radius $r_+$ and angular quantum number $l$ there is an
infinity of QN frequencies (or overtones). 
We shall order them according to
the standard procedure, by increasing imaginary part. Accordingly,
the fundamental QN frequency is defined as the one having the lowest
imaginary part (in absolute value) 
and will be labeled with the integer $n=0$.  The first
overtone has the second lowest imaginary part and is labeled with
$n=1$, and so on. The QN frequencies also have a real part, which 
in general display an increase along with the imaginary part. To the lowest 
value of the imaginary part corresponds the lowest value of 
the real part, to the second lowest 
value of the imaginary part corresponds the second lowest value of 
the real part, and so on. Thus $n$, the overtone number,  
is also a  number that in general 
increases with the real part of the frequency 
(or energy) of the mode. This seems to be a characteristic of
AdS space only, due to the special boundary conditions
associated with this spacetime. 
This, in a sense, is to be expected since the 
wave equation to be studied is a Schr\"odinger type equation, where 
for quantum non-dissipative bound systems, such as the hydrogen atom 
or a particle in an infinite well potential, the 
principal quantum number $n$  (which here has been called the 
overtone number) appears due to the boundary conditions of 
the radial equation, a typical eigenvalue problem, 
and is related directly with the frequency 
of vibration of the orbital. The similarity is not full, though, 
since the boundary condition at the black hole is of a different 
kind. However, for pure AdS spacetimes, when there is no black 
hole and the boundary conditions are of infinite well type, 
the overtone number $n$ is is indeed a principal quantum number 
(see Appendix A).

\subsection{Scalar quasinormal frequencies}
The fundamental scalar QN frequencies were first computed by Horowitz
and Hubeny \cite{horowitz} for intermediate and large black
holes. Konoplya \cite{konoplya} extended these calculations to the
case of small black holes.  Recently Berti and Kokkotas \cite{berti}
rederived all these results.  Here we do for the first time an
extensive search for higher overtones of scalar perturbations. 
Some of the lowest lying modes we find are shown in Tables 1, 2 and 3
for large, intermediate and small black holes, respectively.

\vskip 2mm
\medskip
\noindent {\bf (i) Large black holes -} 
As proven by Horowitz and Hubeny \cite{horowitz} in the large black
hole regime the frequencies must scale as the horizon radius (this can
also be proven easily and directly from the differential equation
(\ref{waveeq})).  
We show in Table 1 the results for a spherically
symmetric mode ($l=0$) for a black hole with $r_+=100$ which is therefore
sufficient to infer the behaviour of all large black holes.  The
fundamental frequency agrees with previous results \cite{horowitz}.
Perhaps the most interesting result in this large black hole regime is
that asymptotically for high overtone number $n$ the frequencies
become evenly spaced and behave like, for $l=0$, 
\begin{equation}
\frac{\omega_{\rm s}}{r_+}=(1.299-2.25 i)n +1.856-2.673i
\,\,,\,\,\,\,(n\,,r_+)\rightarrow \infty \,.  \label{asymptoticscalar}
\end{equation}
Thus the spacing between frequencies is 
\begin{equation}
\frac{ {\omega_{{\rm s}_{\,n+1}}}-{\omega_{{\rm s}_{\,n}}}}{r_+}
=(1.299-2.25 i)
\,\,,\,\,\,\,(n\,,r_+)\rightarrow \infty \,.  \label{asymptoticscalarspacing}
\end{equation}
Moreover, although the offset $1.856-2.673i$ in
(\ref{asymptoticscalar}) is $l$-dependent (this number is different
for $l=1$ scalar perturbations for example), this asymptotic behaviour
for the spacing (\ref{asymptoticscalarspacing}) holds for any value of
$l$.  In fact our search of the QN frequencies for higher values of
$l$ reveal that the results are very very similar to those in Table 1.
We have gone up to $l=4$ for scalar perturbations and the results were
quite insensitive to $l$.
The asymptotic behaviour sets in very quickly as one increases the
mode number $n$.  Typically for $n=10$ equation
(\ref{asymptoticscalar}) already gives a very good approximation.  Indeed,
for $n=10$ we find numerically (see Table 1) 
$\omega_{\rm s}=1486.23753 -2516.90740i$
for a $r_+=100$ black hole, while the asymptotic expression gives
$\omega_{\rm s}=1484.6 - 2517.3i$.

\medskip
\noindent {\bf (ii) Intermediate black holes -}
In Table 2 we show some of the lowest lying scalar QN frequencies
for an intermediate black hole with $r_+=1$. For a black with this size,
one finds again that the spacing does not depend on the angular number $l$
for very high overtone number $n$.
With an error of about 2\% the limiting value for the frequency is, for $l=0$,
\begin{equation}
\omega_{\rm s} \sim (1.97-2.35i)n+2.76-2.7i\,\,,\,\,\,\,n \rightarrow \infty 
\,.
\label{intscalar}
\end{equation}
For QN frequencies belonging to different $l$'s the offset in 
(\ref{intscalar})
is different, but as far as we can tell numerically, not the asymptotic 
spacing implied
by (\ref{intscalar}).
Expression (\ref{intscalar}) for the asymptotic behaviour works well 
again for $n>10$.

\begin{table}
\caption{\label{tab:1} QN frequencies corresponding to $l=0$ scalar
perturbations of a large Schwarzschild-AdS BH ($r_{+}=100$). It can be 
seen that
for large $n$ the modes become evenly spaced. Although not shown here,
our numerical data indicates that this happens for all values of $l$
and also that the spacing is the same, regardless of the value of $l$.
For $l=0$ and for high $n$ the QN frequencies go like 
$\frac{\omega_{\rm s}}{r_+}=(1.299-2.25 i)n +1.856-2.673i$. The corresponding
spacing between consecutive modes seems to be $l$-independent.}
\begin{tabular}{lll|lll}  \hline
$n$  &${\rm Re}[\omega_{QN}]$:&${\rm Im}[\omega_{QN}]:$ &
$n$  &${\rm Re}[\omega_{QN}]$:&${\rm Im}[\omega_{QN}]:$  \\ \hline
0  & 184.95344  & -266.38560 & 7   & 1096.44876 & -1841.88813     \\ \hline
1  & 316.14466  & -491.64354 & 8   & 1226.38317 & -2066.89596      \\ \hline
2  & 446.46153  & -716.75722 & 9   & 1356.31222 & -2291.90222      \\ \hline
3  & 576.55983  & -941.81253 & 10  & 1486.23753 & -2516.90740      \\ \hline
4  & 706.57518  & -1166.8440 & 50  & 6682.78814 & -11516.9823 \\  \hline
5  & 836.55136  & -1391.8641 & 299 & 39030.810  & -67542.308  \\ \hline
6  & 966.50635  & -1616.8779 & 300 & 39160.7272 & -67767.3091 \\ \hline   
\end{tabular}
\end{table}
\begin{table}
\caption{\label{tab:2} QN frequencies corresponding to $l=0$ scalar
perturbations of an intermediate Schwarzschild-AdS BH ($r_+=1$). 
Asymptotically for
large $n$ one finds approximately $\omega_{\rm s} \sim (1.97-2.35i)n+
2.76-2.7i$.} 
\begin{tabular}{lll|lll}  \hline
$n$  &${\rm Re}[\omega_{QN}]$:&${\rm Im}[\omega_{QN}]:$ &
$n$  &${\rm Re}[\omega_{QN}]$:&${\rm Im}[\omega_{QN}]:$   \\ \hline
0  & 2.7982   & -2.6712   & 10 & 22.44671 & -26.20913 \\ \hline
1  & 4.75849  & -5.03757  & 11 & 24.41443   & - 28.55989  \\ \hline
2  & 6.71927  & -7.39449  & 12 & 26.38230   &  -30.91059 \\ \hline
3  & 8.46153  & -9.74852  & 13 & 28.35029   &  -33.26123 \\ \hline
4  & 10.6467 & -12.1012  & 14 & 30.31839   &  -35.61183 \\ \hline
5  & 12.6121 & -14.4533  & 15 & 32.28658   &  -37.96238 \\  \hline
6  & 14.5782 & -16.8049  & 16 & 34.25485   &  -40.31290 \\ \hline
7  & 16.5449 & -19.1562  & 17 & 36.22318   &  -42.66340 \\ \hline
8  & 18.5119 & -21.5073  & 18 & 38.19157   &  -45.01387 \\ \hline
9  & 20.4792 & -23.8583  & 19 & 40.16002   &  -47.36431 \\ \hline
\end{tabular}
\end{table}
\begin{table}
\caption{\label{tab:3} QN frequencies corresponding to $l=0$ scalar
perturbations of a small Schwarzschild-AdS BH ($r_{+}=0.2$). 
Asymptotically for
large $n$ one finds approximately $\omega_{\rm s} \sim (1.69-0.57i)n+
2.29-0.46i$.}
\begin{tabular}{lll|lll}  \hline
$n$  &${\rm Re}[\omega_{QN}]$:&${\rm Im}[\omega_{QN}]:$ &
$n$  &${\rm Re}[\omega_{QN}]$:&${\rm Im}[\omega_{QN}]:$   \\ \hline
0  &  2.47511  & -0.38990 & 6  & 12.45222  & -3.89179 \\ \hline
1  &  4.07086  & -0.98966 & 7  & 14.14065  & -4.46714 \\ \hline
2  &  5.72783  & -1.57600 & 8  & 15.83026  & -5.04186 \\ \hline
3  &  7.40091  & -2.15869 & 9  & 17.52070  & -5.61610  \\  \hline
4  &  9.08118  & -2.73809 & 10 & 19.21191  & -6.18997 \\ \hline
5  &  10.7655  & -3.31557 & 11 & 20.90359  & -6.76355 \\ \hline
\end{tabular}
\end{table}
\vskip 5mm

\medskip 
\noindent {\bf (iii) Small black holes -} 
Our search for
the QN frequencies of small black holes, i.e, black holes with $r_+
\ll 1$ revealed what was expected on physical grounds, and was
uncovered numerically for the first time in \cite{konoplya} for the
fundamental mode: for small black holes, the QN frequencies approach
the frequencies of pure AdS spacetime \cite{burgess,ion} (see
also Appendix A).  In fact we find
\begin{equation} \omega_{\rm s} =2n +l+3 \,\,,r_+\rightarrow 0 \,.
\label{asymptoticscalarsmall} 
\end{equation} 
In Table 3 we show some results for a small black hole with $r_+=0.2$.
We stress that the values presented in Table 3 for the
asymptotic spacing between modes may have an error of about $2 \%$. In
fact it is extremely difficult to find very high overtones of small
black holes, and so it is hard to give a precise extimate of the value
they asymptote to.

\medskip
In summary, we can say that the QN
frequencies tend to be evenly spaced asymptotically as $n$ gets very
large, no matter if the black hole is large, intermediate or small.
Moreover the spacing between consecutive modes is, as far as we can
tell, independent of the angular quantum number $l$.

\subsection{Electromagnetic quasinormal frequencies}
The fundamental electromagnetic QN frequencies were computed for the
first time by Cardoso and Lemos \cite{cardosolemos2}. Recently Berti
and Kokkotas \cite{berti} have redone the calculation showing 
excellent agreement. Here we extend the results to higher overtones.
Some of the lowest lying
electromagnetic frequencies are shown in Tables 4-8.

\medskip
\noindent {\bf (i) Large black holes -} 
As found in \cite{cardosolemos2} large black holes show a
somewhat peculiar behaviour: some of the lowest lying modes have pure
imaginary frequencies, and these are well described by an analytical
formula \cite{cardosolemos2}.  A surprising aspect unveiled for the
first time by the present search is that the number of such modes decreases
as the horizon radius becomes smaller, as can be seen from Tables 4 and 5. 
In other words, for very large black holes the number of imaginary 
modes grows. 
For example, for $r_+=1000$ (Table 4) there are eight pure imaginary modes, 
for $r_+=100$ there are four such modes (see Table 5), and  
for $r_+=10$ there are only two.
If one wants to go for $r_{+}$  larger than $1000$, 
the computation is very time consuming since we use a trial
and error method for finding new modes.  However, not only is this a
completely new piece of data, it also makes us think that infinitely
large black holes may have pure imaginary electromagnetic QN 
frequencies for any overtone number. Perhaps, an infinitly large
black hole cannot vibrate at all. 

Again, we find that for large black holes and $l=1$, the 
frequencies are evenly spaced with 
\begin{equation}
\frac{\omega_{\rm em}}{r_+}=(1.299-2.25 i)n -11.501+12i
\,\,,\,\,\,\,(n\,,r_+)\rightarrow \infty \,.  \label{asymptoticel}
\end{equation}
And a spacing given by 
\begin{equation}
\frac{\omega_{{\rm em}_{\,n+1}}-\omega_{{\rm em}_{\,n}}}{r_+}=(1.299-2.25 i)
\,\,,\,\,\,\,(n\,,r_+)\rightarrow \infty \,.  \label{asymptoticelspacing}
\end{equation}
For different values of the angular quantum number $l$, we find the
same spacing (\ref{asymptoticelspacing}) between consecutive modes,
although the offset in (\ref{asymptoticel}) depends on $l$.
So, asymptotically for large $n$ and large horizon radius the spacing
is the same as for the scalar case!  This is surprising, specially since
the behaviour of the scalar and electromagnetic potentials are
radically different. It is even more surprising the fact that this
asymptotic behaviour does not depend on $l$, as the
electromagnetic potential is strongly $l$-dependent.  Furthermore from
the first electromagnetic overtones one could surely not anticipate
this behaviour.

\begin{table}
\caption{\label{tab:4} QNMs corresponding to $l=1$ electromagnetic
perturbations of a large Schwarzschild-AdS BH ($r_{+}=1000$). 
Notice that now there
are eight pure imaginary modes, still well described by Liu's formula.}
\begin{tabular}{lll|lll}  \hline
$n$  &${\rm Re}[\omega_{QN}]$:&${\rm Im}[\omega_{QN}]:$ &
$n$  &${\rm Re}[\omega_{QN}]$:&${\rm Im}[\omega_{QN}]:$     \\ \hline
0  & 0     & -1500.004789   & 5 &    0      &  -8985.232 \\ \hline
1  & 0     & -2999.982599   & 6 &    0      &  -10596.03 \\ \hline
2  & 0     & -4500.093600   & 7 &    0      &  -11644.76 \\ \hline
3  & 0     & -5999.513176   & 8 & 1219.7    &  -13566.42 \\  \hline
4  & 0     & -7502.69385    & 9 & 2494.6    &  -15847.06 \\ \hline
\end{tabular}
\end{table}
\begin{table}
\caption{\label{tab:5} QNMs corresponding to $l=1$ electromagnetic
perturbations of a large Schwarzschild-AdS BH ($r_{+}=100$). The first 
four modes
are pure imaginary and are well described by Liu's approximation
\cite{cardosolemos2}. For high $n$ the QN freuencies obey, for $l=1$,
$\frac{\omega_{\rm em}}{r_+}=(1.299-2.25 i)n -11.501+12i$. The corresponding
spacing between consecutive modes seems to be $l$-independent.}
\begin{tabular}{lll|lll}  \hline
$n$  &${\rm Re}[\omega_{QN}]$:&${\rm Im}[\omega_{QN}]:$ &
$n$  &${\rm Re}[\omega_{QN}]$:&${\rm Im}[\omega_{QN}]:$     \\ \hline
0  & 0        & -150.0479  & 10 & 799.6171   &  -2171.826 \\ \hline
1  & 0        & -299.8263  & 11 & 927.812    &  -2398.208   \\ \hline
2  & 0        & -450.9458  & 12 & 1056.153   &  -2624.438  \\ \hline
3  & 0        & -595.3691  & 13 & 1184.620   &  -2850.543  \\ \hline
4  & 22.504   & -799.194    & 14 & 1313.192   &  -3076.546  \\  \hline
5  & 162.256  & -1035.098   & 15 & 1441.856   &  -3302.464  \\ \hline
6  & 289.028  & -1263.537   & 16 & 1570.601   &  -3528.310  \\ \hline
7  & 416.247  & -1491.223   & 17 & 1699.416   &  -3754.094  \\ \hline
8  & 543.792  & -1718.409   & 18 & 1828.295   &  -3979.824 \\ \hline
9  & 671.598  & -1945.246   &  19 & 1957.229   &  -4205.508  \\ \hline
\end{tabular}
\end{table}
\begin{table}
\caption{\label{tab:6} QNMs corresponding to $l=1$ electromagnetic
perturbations of an intermediate Schwarzschild-AdS BH ($r_{+}=1$). 
Asymptotically for 
large $n$ the modes become evenly spaced in mode number and behave as 
$\omega_{\rm em} \sim (1.96-2.36i)n+1.45-2.1i$.}
\begin{tabular}{lll|lll}  \hline
$n$  &${\rm Re}[\omega_{QN}]$:&${\rm Im}[\omega_{QN}]:$ &
$n$  &${\rm Re}[\omega_{QN}]$:&${\rm Im}[\omega_{QN}]:$     \\ \hline
0  & 2.163023  & -1.699093  & 10 & 21.067466 & -25.61714  \\ \hline
1  & 3.843819  & -4.151936  & 11 & 23.015470 & -27.98278 \\ \hline
2  & 5.673473  & -6.576456  & 12 & 24.965381 & -30.34713 \\ \hline
3  & 7.553724  & -8.980538  & 13 & 26.916889 & -32.71037 \\ \hline
4  & 9.458385  & -11.37238  & 14 & 28.869756 & -35.07265 \\  \hline
5  & 11.37722  & -13.75633  & 15 & 30.823790 & -37.43413 \\ \hline
6  & 13.30526  & -16.13482  & 16 & 32.778838 & -39.79488\\ \hline
7  & 15.23974  & -18.50933  & 17 & 34.734776 & -42.15499 \\ \hline
8  & 17.17894  & -20.88081  & 18 & 36.691500 & -44.51455 \\ \hline
9  & 19.12177  & -23.24993  & 19 & 38.648922 & -46.87360 \\ \hline
\end{tabular}
\end{table}
\begin{table}
\caption{\label{tab:7} QNMs corresponding to $l=1$ electromagnetic
perturbations of a small Schwarzschild-AdS BH ($r_{+}=0.2$). 
Asymptotically for large $n$ one finds approximately
$\omega_{\rm em} \sim (1.68-0.59i)n+1.87-0.04i$.}
\begin{tabular}{lll|lll}  \hline
$n$  &${\rm Re}[\omega_{QN}]$:&${\rm Im}[\omega_{QN}]:$ &
$n$  &${\rm Re}[\omega_{QN}]$:&${\rm Im}[\omega_{QN}]:$   \\ \hline
0  &  2.63842  & -0.05795 & 6  &  12.00066 & -3.53148  \\ \hline
1  &  3.99070  & -0.47770 & 7  & 13.66436  & -4.12974 \\ \hline
2  &  5.49193  & -1.08951 & 8  & 15.33370  & -4.72479 \\ \hline
3  &  7.07835  & -1.70859 & 9  & 17.00715  & -5.31725 \\  \hline
4  &  8.70165  & -2.32191 & 10 & 18.68370  & -5.90758 \\ \hline
5  &  10.3450  & -2.92920 & 11 & 20.36268  & -6.49615 \\ \hline
\end{tabular}
\end{table}
\begin{table}
\caption{\label{tab:8} The fundamental  ($n=0$) QNMs corresponding to $l=1$
electromagnetic perturbations of a small Schwarzschild-AdS BH 
for several values of $r_{+}$.}
\begin{tabular}{lll|lll}  \hline
$r_{+}$  &${\rm Re}[\omega_{QN}]$:&${\rm Im}[\omega_{QN}]:$ &
$r_{+}$  &${\rm Re}[\omega_{QN}]$:&${\rm Im}[\omega_{QN}]:$     \\ \hline
1/2  & 2.25913  & -0.65731  & 1/10 & 2.85188   &  -0.00064 \\ \hline
1/3  & 2.40171  & -0.29814  & 1/12 & 2.88058   &  -0.00030 \\ \hline
1/4  & 2.53362  & -0.13364  & 1/16 & 2.91363   &  -0.00016 \\ \hline
1/5  & 2.63842  & -0.05795  & 1/18 & 2.92406   &  -0.00009 \\  \hline
1/8  & 2.80442  & -0.00565  & 1/20 & 2.93200   &  -0.00002 \\ \hline
\end{tabular}
\end{table}

\medskip
\noindent {\bf (ii) Intermediate black holes -} 
In Table 6 we show some of the lowest lying electromagnetic QN frequencies
for an intermediate black hole with $r_+=1$. 
For a black with this size,
one finds again that the spacing does not depend on the angular number $l$
for very high overtone number $n$.
With an error of about 2\% the limiting value for the frequency is, for $l=1$,
\begin{equation}
\omega_{\rm em} \sim (1.96-2.36i)n+1.45-2.1i \,\,,\,\,\,\,n 
\rightarrow \infty \,.
\label{intel}
\end{equation}
We note that here too the offset in (\ref{intel}) does depend on $l$, but not
the asymptotic spacing.

\medskip
\noindent {\bf (iii) Small black holes -} 
For small black holes, see Tables 7 and 8, 
the spacing seems also to be equal as for the scalar case, but
since it is very difficult to go very high in mode number $n$ in this
regime, the error associated in estimating the asymptotic behaviour is
higher, and one cannot be completely sure.
Again, the
electromagnetic QN frequencies of very small black holes asymptote to the
pure AdS electromagnetic modes (see Appendix A, where we sketch their
computation).  Indeed we find that \begin{equation} 
\omega_{{\rm em}_{\rm AdS}}
=2n +l+2 \,\,,r_+\rightarrow 0 \,.  \label{asymptoticelsmall}
\end{equation}
This can be clearly seen from Table 8, where we show the fundamental
mode for small black holes of decreasing radius.  As the horizon
radius gets smaller and smaller, the fundamental frequency approaches
the value of $3+\,0\,i$, which is indeed the correct pure AdS mode for $l=1$,
$n=0$, electromagnetic perturbations.  It was conjectured by Horowitz
and Hubeny \cite{horowitz} that for very small black holes in 
AdS space, the imaginary part of the QN frequency for spherically
symmetric perturbations should scale with the horizon area, i.e., with
$r_+^2$.  Their argument was based on a previous result \cite{gibbons}
for the absorption cross section for the $l=0$ component. This conjecture 
was later verified numerically to be correct by Konoplya \cite{konoplya}
for the $l=0$ case.
>From Table 8 it is however apparent that this scaling is no longer
valid for $l=1$ perturbation, and indeed we find it is 
not valid for
$l\neq 0$ perturbations, be it scalar, electromagnetic or gravitational
perturbations.
The reason why the imaginary part no longer scales with the horizon
area for $l\neq 0$ perturbations is due to the fact that the partial
absorption cross section only scales with the horizon area for $l=0$
perturbations. For other $l$'s the behaviour is more complex, and it
could be that there is no simple scaling, or even that the behaviour
is oscillatory with the mass $M$ of the black hole. We refer the
reader to \cite{futterman} for details on the absorption cross section
of black holes.

\vskip 1cm

\subsection{Gravitational quasinormal frequencies}
The fundamental gravitational QN frequencies were computed for the
first time by Cardoso and Lemos \cite{cardosolemos2}.
We remind that there are two sets of gravitational wave equations,
the odd and even ones. Although it was found \cite{cardosolemos2}
that there is a family of the odd modes which is very slowly damped
and pure imaginary, it was possible to prove that for high frequencies
both odd and even perturbations must yield the same QN frequencies.
We present the results for higher overtones of odd perturbations 
in Tables 9-12, and even perturbations in Tables 13-16.

\subsubsection{Odd perturbations}

\medskip
\noindent {\bf (i) Large black holes -} As discussed for the first time
in \cite{cardosolemos2} these exhibit a pure imaginary fundamental mode 
(see Table 9.).
For large black holes, this mode is slowly damped and scales as the inverse
of the horizon radius. Our analysis for higher $l$'s indicates that in the 
large black hole regime an excellent fit to this fundamental pure imaginary mode
is 
\begin{equation}
\omega_{{\rm odd}_{\,n=0}}=-\frac{(l-1)(l+2)}{3r_+}i \,\,,\,\,\,\,r_+\rightarrow \infty. 
\label{pureimaginaryodd}
\end{equation}
This generalizes a previous result by Berti and Kokkotas \cite{berti}
for the $l=2$ case.  
The simplicity of this formula (which is just a fit to our numerical data), 
leads us to believe it is possible to find an analytical explanation for it, 
but such expalnation is still lacking.
In the large black hole regime, asymptotically for high 
overtones one finds,
for $l=2$ for example,
\begin{equation}
\frac{\omega_{\rm odd}}{r_+}=(1.299-2.25 i)n +0.58-0.42i.
\,\,,\,\,\,\,(n\,,r_+)\rightarrow \infty \,.  \label{asymptoticodd}
\end{equation}
This leads to the spacing
\begin{equation}
\frac{\omega_{{\rm odd}_{\,n+1}}-\omega_{{\rm odd}_{\,n}}}{r_+}=(1.299-2.25 i)
\,\,,\,\,\,\,(n\,,r_+)\rightarrow \infty \,,  \label{asymptoticspacingodd}
\end{equation}
which, as our results indicate is again $l$-independent.
Again, the offset in (\ref{asymptoticodd}) depends on $l$.

\medskip
\noindent {\bf (ii) Intermediate black holes -}
Results for the odd QN frequencies of an intermediate ($r_+=1$) 
black hole are shown in Table 10. 
With an error of about 5\% the limiting value for the frequency is, for $l=2$,
\begin{equation}
\omega_{\rm odd} \sim (1.97-2.35i)n+0.93-0.32i \,\,,\,\,\,\,n \rightarrow \infty \,.
\label{intodd}
\end{equation}
We note that here too the offset in (\ref{intodd}) does depend on $l$, but not
the asymptotic spacing, with a numerical error of about 5\%.

\medskip
\noindent {\bf (iii) Small black holes -} 
The behavior for small black holes is shown in Tables 11 and 12.
As the black hole gets smaller, the pure imaginary mode gets more
damped: the imaginary part increases, as can be seen from Table 12,
where we show the two lowest QN frequencies for small black holes with
decreasing radius.  As mentioned by Berti and Kokkotas \cite{berti}
the ordering of the modes here should be different.  However, since
one can clearly distinguish this pure imaginary mode as belonging to a
special family, we shall continue to label it with $n=0$.  We have not
been able to follow this mode for black holes with $r_+ <0.5$, and so
Table 12 does not show any pure imaginary modes for horizon radius
smaller than $0.5$.  We note that, as for the scalar and
electromagnetic cases, here too the modes are evenly spaced, with a
spacing which seems to be independent of $l$ no matter if the black
hole is large or small.  For very small black holes, the frequencies
reduce to their pure AdS values, computed in Appendix A, to
wit
\begin{equation} 
\omega_{\rm odd}
=2n +l+2 \,,\;\;\;r_+\rightarrow 0 \,. 
\label{asymptoticoddsmall}
\end{equation}
One can see this more clearly from Table 12, where in fact for very 
small black holes
the frequency rapidly approaches (\ref{asymptoticoddsmall}).
Again, for small black holes, the imaginary part does not scale with 
the horizon area, 
by the reasons explained before, in section 3.2.

\begin{table}
\caption{\label{tab:9} QNMs corresponding to $l=2$ odd gravitational 
perturbations of a large Schwarzschild-AdS BH ($r_{+}=100$). The 
fundamental QN frequency
is pure imaginary and seems to be well described by the formula 
$\omega_{n=0}=\frac{-(l-1)(l+2)}{3r_+}i$ valid only in the large black hole regime.
In the large $n$ limit one finds 
$\frac{\omega_{\rm odd}}{r_+}=(1.299-2.25 i)n +0.58-0.42i$. The corresponding
spacing between consecutive modes seems to be $l$-independent.}
\begin{tabular}{lll|lll}  \hline
$n$  &${\rm Re}[\omega_{QN}]$:&${\rm Im}[\omega_{QN}]:$ &
$n$  &${\rm Re}[\omega_{QN}]$:&${\rm Im}[\omega_{QN}]:$ \\ \hline
0  & 0          & -  0.013255  & 6  & 836.55392    & -1391.86345 \\ \hline
1  & 184.95898  & - 266.38403  & 7   &  966.50872  & -1616.87735 \\ \hline
2  & 316.14887  & - 491.64242  & 8   &  1096.45098 & -1841.88755 \\ \hline
3  & 446.46505  & - 716.75629  & 9   &  1226.38527 & -2066.89540 \\ \hline
4  & 576.56293  & - 941.81172  & 10  &  1356.31422 & -2291.90170 \\  \hline
5  & 706.57797  & - 1166.8433  & 50  &  6552.87704 & -11291.9807  \\ \hline
\end{tabular}
\end{table}
\begin{table}
\caption{\label{tab:10} QNMs corresponding to $l=2$ odd gravitational
perturbations of an intermediate Schwarzschild-AdS BH ($r_{+}=1$). Asymptotically
for large $n$ one finds approximately
$\omega_{\rm odd} \sim (1.97-2.35i)n+0.93-0.32i$.}
\begin{tabular}{lll|lll}  \hline
$n$  &${\rm Re}[\omega_{QN}]$:&${\rm Im}[\omega_{QN}]:$ &
$n$  &${\rm Re}[\omega_{QN}]$:&${\rm Im}[\omega_{QN}]:$   \\ \hline
0  & 0         & -2         & 10 & 20.604949 & -23.803860 \\ \hline
1  & 3.033114  & -2.404234  & 11 & 22.567854   &  -26.157246 \\ \hline
2  & 4.960729  & -4.898194  & 12 & 24.531429   &  -28.510214 \\ \hline
3  & 6.905358  & -7.289727  & 13 & 26.495564   &  -30.862849  \\ \hline
4  & 8.854700  & -9.660424  & 14 & 28.460169   &  -33.215214 \\  \hline
5  & 10.80784  & -12.02344  & 15 & 30.425175   &  -35.567355 \\ \hline
6  & 12.76384  & -14.38266  & 16 & 32.390524   &  -37.919308 \\ \hline
7  & 14.72199  & -16.73969  & 17 & 34.356173   &  -40.271103  \\ \hline
8  & 16.68179  & -19.09530  & 18 & 36.322082   &  -42.622761 \\ \hline
9  & 18.64286  & -21.44994  & 19 & 38.288221   &  -44.974301 \\ \hline
\end{tabular}
\end{table}
\begin{table}
\caption{\label{tab:11} QNMs corresponding to $l=2$ odd gravitational
perturbations of a small Schwarzschild-AdS BH ($r_{+}=0.2$). Asymptotically for
large $n$ one finds approximately $\omega_{\rm odd} \sim (1.69-0.59i)n+2.49+0.06i$.}
\begin{tabular}{lll|lll}  \hline
$n$  &${\rm Re}[\omega_{QN}]$:&${\rm Im}[\omega_{QN}]:$ &
$n$  &${\rm Re}[\omega_{QN}]$:&${\rm Im}[\omega_{QN}]:$   \\ \hline
0  &  2.404    & -3.033   & 6  &  12.67161  & -3.43609 \\ \hline
1  &  4.91594  & -0.30408 & 7  &  14.33020  & -4.05366 \\ \hline
2  &  6.30329  & -0.89773 & 8  &  15.99881  & -4.66448 \\ \hline
3  &  7.82330  & -1.53726 & 9  &  17.67433  & -5.26955 \\  \hline
4  &  9.40720  & -2.17744 & 10 &  19.35465  & -5.86978\\ \hline
5  &  11.0279  & -2.81083 & 11 &  21.03839  & -6.46596  \\ \hline
\end{tabular}
\end{table}
\begin{table}
\caption{\label{tab:12} The fundamental ($n=0$) QNMs corresponding to $l=2$
odd gravitational perturbations of a small Schwarzschild-AdS BH 
for several values of $r_{+}$.}
\begin{tabular}{lll|lll}  \hline
$r_{+}$  &${\rm Re}[\omega_{QN}]$:&${\rm Im}[\omega_{QN}]:$ &
$r_{+}$  &${\rm Re}[\omega_{QN}]$:&${\rm Im}[\omega_{QN}]:$     \\ \hline
0.8 $(n=0)$ &  0              & -3.045373 & 0.5 $(n=1)$ & 3.03759  &  -0.71818\\ \hline
0.8 $(n=1)$ &  2.89739        & -1.69556  & 0.4         & 3.16209   & -0.43092\\\hline
0.7 $(n=0)$ & 0               & -3.83538  & 0.3         & 3.35487   & -0.17320\\ \hline
0.7 $(n=1)$ &  2.90665        & -1.34656  & 0.2         & 3.62697   & -0.01792\\ \hline
0.6 $(n=0)$ & 0               & -4.901973 & 0.1         & 3.84839   & -0.00005\\ \hline
0.6 $(n=1)$ & 2.95550         & -1.02196  & 1/15        & 3.90328   & -0.00001\\  \hline
0.5 $(n=0)$ & 0               & -6.40000  & 1/20        & 3.92882   & -0.000002\\ \hline
\end{tabular}
\end{table}

\medskip
In conclusion,
the higher overtones of odd perturbations follow a pattern very
similar to the scalar case. We note that the asymptotic behaviour
sets in very quickly, much like what happened for scalar and electromagnetic
perturbations. Typically the formulas yielding the asymptotic behaviour
work quite well for $n>10$.
We are now
able to prove that for sufficiently high
frequencies the scalar and gravitational perturbations are
isospectral, a mystery that remained in 
\cite{cardosolemos2},  This is done in section 4.1 below.

\subsubsection{Even perturbations}

Let us now briefly discuss the even modes.  As found previously
\cite{cardosolemos2} these modes behave very similar to the scalar
ones.  Yet, the even gravitational modes are stipulated by a more
complicated potential, and we have to truncate the series in power of
$x-x_{+}$ at larger $N$, which makes the whole procedure be more time
consuming. That is why when considering small black holes 
we were restricted only by first seven modes in that case. It is,
however, sufficient to see that even gravitational QNMs,
similar to other kind of perturbations, tend to arrange into
equidistant spectrum under the increasing of $n$.
We show in Tables 13-16 the numerical results for the QN frequencies
of even gravitational perturbations.

\medskip
\noindent {\bf (i) Large black holes -}
Results for the QN frequencies of large black holes are shown in
Table 13.
In this regime one finds for $l=2$ even perturbations
\begin{equation}
\frac{\omega_{\rm even}}{r_+}=(1.299-2.25 i)n +1.88-2.66i.
\,\,,\,\,\,\,(n\,,r_+)\rightarrow \infty \,,  \label{asymptoticeven}
\end{equation}
leading to the spacing
\begin{equation}
\frac{\omega_{{\rm even}_{\,n+1}}-
\omega_{{\rm even}_{\,n}}}{r_+}=(1.299-2.25 i)
\,\,,\,\,\,\,(n\,,r_+)\rightarrow \infty \,,  \label{asymptoticspacingeven}
\end{equation}
which once more turns out to be $l$-independent!  All the results
concerning the spacing of frequencies for large black holes have a
very good precision, since in this regime it is possible to go very
far out in overtone number (typically $n=300$ is enough to achieve a
0.1\% accuracy for the spacing).

\medskip
\noindent {\bf (ii) Intermediate black holes -}
In Table 14 we show some of the lowest lying even gravitational QN frequencies
for an intermediate black hole with $r_+=1$. 
For a black with this size,
one finds again that the spacing does not seem to depend on the angular number $l$
for very high overtone number $n$.
With an error of about 5\% the limiting value for the frequency is, for $l=2$,
\begin{equation}
\omega_{\rm even} \sim (1.96-2.35i)n+2.01-1.5i \,\,,\,\,\,\,n \rightarrow \infty \,.
\label{inteven}
\end{equation}
We note that here too the offset in (\ref{intel}) does depend on $l$, but not
the asymptotic spacing.

\begin{table}
\caption{\label{tab:13} QNMs corresponding to $l=2$ even gravitational 
perturbations of a large Schwarzschild-AdS BH ($r_{+}=100$). For large $n$, one finds
$\frac{\omega_{\rm even}}{r_+}=(1.299-2.25 i)n +0.58-0.42i$. The corresponding
spacing between consecutive modes seems to be $l$-independent.}
\begin{tabular}{lll|lll}  \hline
$n$  &${\rm Re}[\omega_{QN}]$:&${\rm Im}[\omega_{QN}]:$ &
$n$  &${\rm Re}[\omega_{QN}]$:&${\rm Im}[\omega_{QN}]:$ \\ \hline
0  & 184.97400 & -266.351393  & 6   & 966.609780 & -1616.695872  \\ \hline
1  & 316.17838 & -491.584999  & 7   & 1096.56635 & -1841.681256 \\ \hline
2  & 446.50884 & -716.674054  & 8   & 1226.51495 & -2066.664293 \\ \hline
3  & 576.62103 & -941.70468   & 9   & 1356.45821 & -2291.645761 \\ \hline
4  & 706.65039 & -1166.71147  & 10  & 1486.39776 & -2516.626168 \\  \hline
5  & 836.64066 & -1391.70679  & 50  & 6683.51993 & -11515.70869 \\ \hline
\end{tabular}
\end{table}
\begin{table}
\caption{\label{tab:14} QNMs corresponding to $l=2$ even gravitational
perturbations of an intermediate Schwarzschild-AdS BH ($r_{+}=1$). Asymptotically
for large $n$ one finds approximately
$\omega_{\rm even} \sim (1.96-2.35i)n+2.01-1.5i$.}
\begin{tabular}{lll|lll}  \hline
$n$  &${\rm Re}[\omega_{QN}]$:&${\rm Im}[\omega_{QN}]:$ &
$n$  &${\rm Re}[\omega_{QN}]$:&${\rm Im}[\omega_{QN}]:$   \\ \hline
0  & 3.017795  & -1.583879  & 10 & 21.68949   &  -24.98271  \\ \hline
1  & 4.559333  & -3.810220  & 11 & 23.64402   &  -27.33549 \\ \hline
2  & 6.318337  & -6.146587  & 12 & 25.60052   &  -29.68799 \\ \hline
3  & 8.168524  & -8.500194  & 13 & 27.55860   &  -32.04026   \\ \hline
4  & 10.061220 & -10.85631  & 14 & 29.51796   &  -34.39234  \\  \hline
5  & 11.976813 & -13.21224  & 15 & 31.47838   &  -36.74424  \\ \hline
6  & 13.906140 & -15.56749  & 16 & 33.43969   &  -39.09600 \\ \hline
7  & 15.844371 & -17.92208  & 17 & 35.40174   &  -41.44762  \\ \hline
8  & 17.788721 & -20.27609  & 18 & 37.36444   &  -43.79914  \\ \hline
9  & 19.737469 & -22.62960  & 19 & 39.32769   &  -46.15057\\ \hline
\end{tabular}
\end{table}
\begin{table}
\caption{\label{tab:15} QNMs corresponding to $l=2$ even gravitational
perturbations of a small Schwarzschild-AdS BH ($r_{+}=0.2$). Asymptotically for
large $n$ one finds approximately $\omega_{\rm even} \sim (1.61-0.6i)n+2.7+0.37i$.}
\begin{tabular}{lll|lll}  \hline
$n$  &${\rm Re}[\omega_{QN}]$:&${\rm Im}[\omega_{QN}]:$ &
$n$  &${\rm Re}[\omega_{QN}]$:&${\rm Im}[\omega_{QN}]:$   \\ \hline
0  & 3.56571  & -0.01432 & 3  & 7.65872  & -1.42994 \\ \hline
1  & 4.83170  & -0.26470 & 4  &  9.20424 & -2.04345 \\ \hline
2  & 6.17832  & -0.82063 & 5  & 10.78800 & -2.65360 \\ \hline
\end{tabular}
\end{table}
\begin{table}
\caption{\label{tab:16} The fundamental ($n=0$) QNMs corresponding to $l=2$
even gravitational perturbations of a small Schwarzschild-AdS BH 
for several values of $r_{+}$.}
\begin{tabular}{lll|lll}  \hline
$r_{+}$  &${\rm Re}[\omega_{QN}]$:&${\rm Im}[\omega_{QN}]:$ &
$r_{+}$  &${\rm Re}[\omega_{QN}]$:&${\rm Im}[\omega_{QN}]:$       \\ \hline
0.8  & 2.91541  & -1.18894  & 0.3 & 3.29299   &  -0.14103       \\ \hline
0.7  & 2.90591  & -0.98953  & 0.2 & 3.56571   &  -0.01432       \\ \hline
0.6  & 2.92854  & -0.78438  & 0.1 & 3.80611   &  -0.00005       \\ \hline
0.5  & 2.98985  & -0.57089  &1/15 & 3.8735    &  -0.00001      \\  \hline
0.4  & 3.10317  & -0.35043  &1/20 & 3.90852   &  -0.000002       \\ \hline
\end{tabular}
\end{table}
\medskip
\noindent {\bf (iii) Small black holes -}
The behavior for small black holes is shown in Tables 15 and 16.
Our search for
the QN frequencies of small black holes, i.e, black holes with $r_+
\ll 1$ revealed again what was expected on physical grounds: 
for small black holes, the QN frequencies approach
the frequencies of pure AdS spacetime (see Appendix A).  
In fact we find
\begin{equation} \omega_{{\rm even}_{\rm AdS}} =2n +l+2 \,\,,r_+\rightarrow 0 \,.
\label{asymptoticevensmall} 
\end{equation} 
In Table 15 we show the lowest lying QN frequencies for a small black
hole ($r_+=0.2$).  We stress that the values presented in Table 15 (as
a matter of fact, all the Tables containing data for small black
holes) for the asymptotic spacing between modes may have an error of
about $2 \%$. In fact it is extremely difficult to find very high
overtones of small black holes, and so it is hard to give a precise
extimate of the value they asymptote to.
In Table 16 we show some the fundamental even QN frequencies for  
small black holes of decreasing radius, and one can clearly see how
the fundamental frequency approaches the pure AdS value given
in Appendix A.

\section{Discussion of the results}
\subsection{Why are the scalar and gravitational perturbations
isospectral in the large black hole regime?}
In a previous paper (section IIIC in \cite{cardosolemos2}), we have
showed why the odd and even gravitational perturbations yield the same
QN frequencies for large frequencies.  The whole approach was
based on the fact that the odd and even gravitational potentials are
superpartner potentials \cite{cooper}, i.e., they are related to one
another via
$
V_{\rm odd}=W^2 {+} \frac{dW}{dr_*} +\beta \,,
\quad
V_{\rm even}=W^2 {-} \frac{dW}{dr_*} +\beta \,,
$
where $\beta=-\frac{\alpha^2+2\alpha^3+\alpha^4}{9M^2}$. The function $W$ is
$
W= \frac{2M}{r^2}+\frac{-3-2\alpha}{3r}+\frac{3\alpha^2+2\alpha^2+27M^2}
{3\alpha\left(3M+\alpha \,r\right)}-
\frac{1}{3}\left(\frac{\alpha}{M}+\frac{\alpha^2}{M}+\frac{9M}{\alpha}\right)$.
For more details we refer the reader to \cite{cardosolemos2}.
We shall now see that a similar method can be applied to show that in the
large black hole regime, scalar and gravitational perturbations
are isospectral for large QN frequencies.
To begin with, we note that the potentials $V1$ and $V2$ defined by 
\begin{equation}
V1={\tilde f}  \left( \frac{2}{R^2}+\frac{2M}{r^3} \right) \,,
\label{V1}
\end{equation}
and
\begin{equation}
V2={\tilde f} \left(\frac{a}{r^2}-\frac{6M}{r^3} \right)\,,
\label{V2}
\end{equation}
with $\tilde{f}=\frac{r^2}{R^2}+\frac{a}{2}-\frac{2M}{r}$, 
and $a$ any constant, 
are superpartner potentials.
The superpotential $\tilde {W}$ is in this case is given by 
\begin{equation}
\tilde {W}=\frac{r}{R^2}+\frac{a}{2r}-\frac{2M}{r^2}\,.
\label{Wtilde}
\end{equation}
Thus, the two superpartner potentials $V1$ and $V2$ can be expressed 
in terms of $\tilde{W}$ as
\begin{equation}
V1=\tilde{W}^2 {+} \frac{d\tilde{W}}{dr_*}  \,,
\quad
V2=\tilde{W}^2 {-} \frac{d\tilde{W}}{dr_*} \,.
\label{V12}
\end{equation}
Why are these two potentials of any interest?  Because in the
large $r_+$ limit, which we shall take to be $r_+ \gg a$, 
we have $\tilde {f} \sim
r^2-\frac{2M}{r}$. Notice now that in this large $r_+$ limit the scalar
potential (\ref{Vscalar}) is $V_{\rm s}\sim f(2+\frac{2M}{r^3})$, with
$f\sim r^2-\frac{2M}{r}$, since in this limit and with $r_+ \gg l$,
one has $\frac{l(l+1)}{r^2}\ll 2$. Thus, $V1$ reduces to the
scalar potential and $V2$ to the gravitational odd potential, provided
we take $a=l(l+1)$.  It then follows from the analysis in
\cite{cardosolemos2}  (section IIIC) that for large black holes these
two potentials should yield the same frequencies.

\subsection{Future Directions}
The preceding sections have shown that the QNMs of
Schwarzschild-AdS black holes have a universal behavior in
the asymptotic regime of high overtones.  This was verified explicitly
and with great accuracy for the large black hole regime, where we
showed numerically that the spacing does not depend on the
perturbation in question and is equal to 
\begin{equation}
\frac{\omega_{n+1}-\omega_{n}}{r_+}=(1.299-2.25 i)
\,\,,\,\,\,\,(n\,,r_+)\rightarrow \infty \,.
\label{asymptoticelspacingUniv} \end{equation} 
We conjecture that the asymptotic behavior is the same for all kinds
of perturbations irrespectively of the black hole size, i.e., a fixed
horizon radius $r_+$ Schwarzschild-AdS black hole will have
an asymptotic spacing between consecutive QN frequencies which is the
same for scalar, electromagnetic and gravitational perturbations. The
difficulty in extracting very high overtones for small black holes
however, prevents us from having an irrefutable numerical proof of
this.  It would be extremely valuable to have some kind of analytical
scheme for extracting the asymptotic behavior, much as has been done
for the asymptotically flat space by Motl and Neitzke
\cite{hoddreyerkunstattercorichimotldreyer2}.  However it looks quite
difficult to make any analytical approximation in asymptotically
AdS spaces, although there have been some atempts at this
recently (see for example Musiri and Siopsis \cite{qn}).  
We also note that the spacing
(\ref{asymptoticelspacingUniv}) was already found to be true by Berti
and Kokkotas \cite{berti} for the scalar and gravitational
cases for the lowest radiatable multipole, i.e, $l=0$ and $l=2$
scalar and gravitational perturbations respectively.  We have
concluded that, surprisingly, the spacing (\ref{asymptoticelspacingUniv})
also works for electromagnetic case and for any value of $l$.
It was observed that,
despite having such different potentials the scalar, the
electromagnetic and gravitational QN frequencies have the same
asymptotic behavior.  Can one formulate some very general conditions
the potentials should obey in order to have the same asymptotic
solutions? This is still an open question.

There has been recently an exciting development trying to relate the
asymptotic QN frequencies with the Barbero-Immirzi parameter
\cite{hoddreyerkunstattercorichimotldreyer2,baez}.  In fact it was
observed, in the Schwarzschild case, that asymptotically for high
overtones, the real part of the QN frequencies was a constant,
$l$-independent, and using some (not very clear yet) correspondence
between classical and quantum states, was just the right constant to
make Loop Quantum Gravity give the correct result for the black hole
entropy.  Of course it is only natural to ask whether such kind of
numerical coincidence holds for other spacetimes.  We have seen that
apparently we are facing, in AdS space, a universal behavior, i.e.,
the asymptotic QN frequencies do not depend on the kind of
perturbations, and also don't depend on $l$.  However, and in contrast
with asymptotically flat space, the real part of the asymptotic QN
frequency is not a constant, but rather increases linearly with the
mode number $n$.  This is no reason to throw off the initial
motivation of seeking some kind of relation between Loop Quantum
Gravity and QNMs, after all, there are no predictions for AdS space.  

Finally we point out that the asymptotic behavior studied here for the 
Schwarzschild-AdS black hole will hold also for other black holes in 
asymptotically AdS. One example of these is the 
black hole
with non-trivial topology \cite{lemos}. 
The general line element for this spacetime is
\cite{lemos}: 
\begin{equation}
ds^{2}= f(r)\,dt^{2}- f(r)^{-1}dr^{2}-r^2\left( d\theta^{2}
+d\phi^{2}\right)\,
\label{lineelementlemos}
\end{equation}
where 
\begin{equation}
f(r)=\frac{
r^2}{R^2}-\frac{4MR}{r}\, ,
\label{f(r)}
\end{equation}
where $M$ is the ADM mass of the black hole, and $R$ is the AdS
radius. There is a horizon at $r_+=(4M)^{1/3}R$.  The range of the
coordinates $\theta$ and $\phi$ dictates the topology of the black hole
spacetime.  For a black hole with toroidal topology, a toroidal black
hole, the coordinate $\theta$  ranges from
$0$ to $2\pi$, and $\phi$ ranges from $0$ to $2\pi$ as well.  For the
cylindrical black hole, or black string, the coordinate $\theta$ has range
$-\infty<R\,\theta<\infty$, and $0\leq \phi <2\pi$. For the planar black hole,
or black membrane, the coordinate $\phi$ is further decompactified
$-\infty<R\,\phi<\infty$ \cite{lemos}.  The fundamental QN frequencies
for these black holes were computed in \cite{cardosolemos3}, where it
was verified that they follow the same pattern as for
Schwarzschild-AdS black holes. Indeed one easily sees that in the
large black hole regime they both should yield the same results as the
potentials are equal in this regime (compare the potentials in
\cite{cardosolemos3} with the ones in the present work).  In
particular the asymptotic behavior will be the same.

\section{Conclusion}
We have done an extensive search for higher overtones $n$ of the
QNMs of Schwarzschild-AdS BH corresponding to scalar,
electromagnetic, and gravitational perturbations.
We have shown that:
(i) No matter what the size of the black hole is, the QN frequencies
are evenly spaced, both in the real and in the imaginary component,
for high overtone number $n$; 
(ii) The spacing between consecutive modes is independent of the perturbation.
This means that scalar, electromagnetic and gravitational perturbations all
have, asymptotically, the same spacing between modes. This is one of the major
findings in this work, together with the fact that this spacing seems to
be also independent of the angular quantum number $l$;
(iii) We were able to prove that the scalar and gravitational QN frequencies
must asymptotically be the same;
(iv) The electromagnetic QN frequencies of large black
holes have a number of first overtones with pure imaginary parts, and
the higher the black hole radius $r_{+}$, the higher the number of these first
pure damped, non-oscillating modes;
(v) Finally, we have computed analytically the electromagnetic and
gravitational pure AdS modes, and we have shown numerically that the
QN frequencies of very small black holes asymptote to these pure AdS
modes;

\section*{Acknowledgements}
It is a pleasure to acknowledge stimulating correspondence related to this
problem with Veronica Hubeny.
This work
was partially funded by Funda\c c\~ao para a Ci\^encia e Tecnologia
(FCT) -- Portugal through project PESO/PRO/2000/4014. VC  also
acknowledges financial support from FCT through PRAXIS XXI programme.
JPSL acknowledges finantial support from ICCTI/FCT and 
thanks Observat\'orio Nacional do Rio de Janeiro for
hospitality.

\vskip 2cm
\appendix
\section{Pure AdS Normal Modes for Electromagnetic and 
Gravitational Perturbations}
\label{apendice}
In this appendix we shall briefly outline how to compute the pure
modes of AdS space (no black hole, $M=0$) for
electromagnetic and gravitational perturbations. The scalar case was
dealt with by Burgess and Lutken \cite{burgess} (see also \cite{ion}).  
In pure AdS space the electromagnetic
and gravitational potentials (both odd and even) are
\begin{equation}
V=\left(\frac{r^2}{R^2}+1\right)\frac{l(l+1)}{r^2}\,,
\label{potpureAdS}
\end{equation}
as can be seen by substituting $M=0$ in (\ref{Velectromagnetic})-
(\ref{veven}).
Also in this case the relation $r(r_*)$ takes the simple form
\begin{equation}
r=R\tan{\frac{r_*}{R}}\,,
\label{rrtort}
\end{equation}
and therefore the potential (\ref{potpureAdS}) takes a simple form in the
$r_*$ coordinate, namely
\begin{equation}
V=\frac{l(l+1)}{R^2\sin\left({\frac{r_*}{R}}\right)^2}\,.
\label{potpureAdSrtort}
\end{equation}
To proceed, we note that the change of variable 
$x=\sin\left({\frac{r_*}{R}}\right)^2$
leads the wave equation to a hypergeometric equation,
\begin{equation}
\frac{\partial^{2} \Psi(x)}{\partial x^{2}} +
\frac{\tilde{\tau}}{\sigma}\frac{\partial \Psi(x)}{\partial x} +
\frac{\tilde{\sigma}}{\sigma ^2}\Psi(x)=0 \,,
\label{waveeqhyperg}
\end{equation}
with
\begin{eqnarray}
\tilde{\sigma}= 4(\omega R)^2 x(1-x) -4l(l+1)(1-x)\,,\\
\label{sigma}
\sigma=4x(1-x)\,,\\
\label{ttau}
\tilde{\tau}=2(1-2x)\,.
\end{eqnarray}
To put this in a more standard form, one changes wavefunction
by defining
\begin{equation}
\Psi(x)=\sqrt{x-1}\;x^{\frac{l+1}{2}}\;Z(x)\,,
\label{chgwavefunction}
\end{equation}
and one gets the following standard hypergeometric
differential equation for $Z$:
\begin{equation}
\sigma \frac{\partial^{2} Z(x)}{\partial x^{2}} +
\tau \frac{\partial Z(x)}{\partial x} +
\lambda Z(x)=0 \,,
\label{waveeqhyperg2}
\end{equation}
with $\sigma$ defined in (\ref{sigma}) and
\begin{eqnarray}
\tau=6-4l(x-1)-12x\,, \\
\lambda=-4-4l-l^2+\omega ^2 \,.
\label{hypergparameters2}
\end{eqnarray}
By requiring well behaved fields everywhere a simple analysis
\cite{uvarov} then shows that the following constraint needs
to be satisfied,
\begin{equation}
\omega \,R=2n+l+2 \,.
\label{frequenciespureadselectrgrav}
\end{equation}
These are the pure AdS frequencies for electromagnetic and gravitational
perturbations, corresponding to pure AdS
normal modes of the corresponding fields.
One can compare the frequencies in (\ref{frequenciespureadselectrgrav}) 
with the scalar frequencies corresponding to pure
AdS modes \cite{burgess,ion},
$
\omega_{\rm s}\, R=2n+l+3 \,.
$

\vskip 2cm


\begin{thebibliography}{99}

\bibitem{reggeW} T. Regge, J. A. Wheeler, Phys. Rev. {\bf 108},
1063 (1957).

\bibitem{Chandra1} S. Chandrasekhar, and S. Detweiler,
 Proc. R. Soc. London, Ser. A {\bf 344}, 441 (1975).

\bibitem{kokkotasnollert1}  K. D. Kokkotas and B. G. Schmidt,
Living Rev. Rel. {\bf2}, 2 (1999);
H. P. Nollert, Class. Quantum Grav.
{\bf 16}, R159 (2000).

\bibitem{hoddreyerkunstattercorichimotldreyer2}  S. Hod,
Phys. Rev. Lett. {\bf 81}, 4293 (1998);
O. Dreyer, Phys. Rev. Lett. {\bf 90}, 081301 (2003);
G. Kunstatter, gr-qc/0212014;
A. Corichi, gr-qc/0212126;
L. Motl, gr-qc/0212096;
L. Motl, A. Neitzke, hep-th/0301173;
A. Maassen van den Brink, gr-qc/0303095.


\bibitem{baez} J. Baez, in {\it Matters of gravity},
p. 12, ed. J. Pullin, gr-qc/0303027.

\bibitem{kokoko} E. Berti and K. D. Kokkotas, hep-th/0303029;
R. A. Konoplya, gr-qc/0303052;


\bibitem{konoplya2}
R. A. Konoplya, Phys. Lett. B {\bf{ 550}}, 117 (2002); 
V. Ferrari, M. Pauri and
F. Piazza, Phys. Rev. D {\bf 63} 064009 (2001);
R. A. Konoplya, Gen. Relativ. Grav. 
{\bf{34}}, 329 (2002).

\bibitem{mm1} F. Mellor and I. G. Moss,
Phys. Rev. D {\bf 41}, 403 (1990); I. G. Moss and J. P. Norman,
Class. Quant. Grav. {\bf 19}, 2323-2332 (2002).

\bibitem{cardosolemos1}  V. Cardoso and J. P. S. Lemos,
Phys. Rev. D {\bf 67}, 084020 (2003).

\bibitem{molina} C. Molina, gr-qc/0304053.

\bibitem{brink} A. Maassen van den Brink, gr-qc/0304092.

\bibitem{suneeta} V. Suneeta, gr-qc/0303114.

\bibitem{maldacena} J. Maldacena, Adv. Theor. Math. Phys.
{\bf 2}, 253(1998).

\bibitem{qn}
S. F. J. Chan and R. B. Mann,
Phys. Rev. D {\bf 55}, 7546 (1997);
V. Cardoso and J. P. S. Lemos,
Phys. Rev. D {\bf 63}, 124015 (2001);
D. Birmingham, I. Sachs and S. N. Solodukhin,
Phys. Rev. Lett. {\bf 88}, 151301 (2002); hep-th/0212308;
B. Wang, C. Molina and E. Abdalla, 
Phys. Rev.  D {\bf 63}, 084001 (2001);
W. T. Kim and J. J. Oh, Phys. Lett. B {\bf 514}, 155 (2001);
R. A. Konoplya, Phys. Rev.  D {\bf 66}, 084007 (2002);
A. O. Starinets, Phys. Rev. D {\bf 66}, 124013 (2002);
Y. Kurita and M. Sakagami, Phys. Rev. D {\bf 67}, 024003 (2003);
R. Aros, C. Martinez, R. Troncoso and J. Zanelli,
Phys. Rev. D {\bf 67}, 044014 (2003);
D. T. Son, A. O. Starinets,
J.H.E.P. {\bf 0209}:042, (2002);
V. Cardoso, O. J. C. Dias and J. P. S. Lemos, 
Phys. Rev. D {\bf 67}, 064026 (2003);
S. Musiri and G. Siopsis, hep-th/0301081.

\bibitem{horowitz} G. T. Horowitz and V. E. Hubeny, 
Phys. Rev. D {\bf 62}, 024027 (2000).

\bibitem{konoplya} R. A. Konoplya, 
Phys. Rev. D {\bf 66}, 044009 (2002).

\bibitem{cardosolemos2} V. Cardoso and J. P. S. Lemos, 
Phys. Rev. D {\bf 64}, 084017 (2001).

\bibitem{berti} E. Berti and K. D. Kokkotas, 
Phys. Rev. D {\bf 67}, 064020 (2003).

\bibitem{lemos} J. P. S. Lemos, Class. Quantum Grav. {\bf 12}, 1081
(1995); Phys. Lett. B {\bf 353}, 46 (1995); J. P. S. Lemos and
V. T. Zanchin, Phys. Rev. D {\bf 54}, 3840 (1996); see also the review
paper: J. P. S. Lemos, {\it Black holes with toroidal, cylindrical and
planar horizons in anti-de Sitter spacetimes in general relativity and
their properties}, ``Recent developments in astronomy and 
astropysics'', Proceedings of the 10th Portuguese Meeting on
Astronomy and Astrophysics, edited by J. P. S. Lemos et al (World
Scientific, 2001, gr-qc/0011092.


\bibitem{cardosolemos3} V. Cardoso and J. P. S. Lemos,
Class. Quant. Grav. {\bf 18}, 5257 (2001).

\bibitem{burgess}
C. P. Burgess and C. A. Lutken,
Phys. Lett. B {\bf 153}, 137(1985). 

\bibitem{ion} I. I. Cotaescu, 
Phys. Rev. D {\bf 60}, 107504 (1999).

\bibitem{gibbons} S. R. Das, G. Gibbons and S. D. Mathur,
Phys. Rev. Lett. {\bf 78}, 417(1997).
       
\bibitem{futterman} J. A. H. Futterman, F. A. Handler and R. A. Matzner,
{\it Scattering from Black Holes},
(Cambridge University Press, Cambridge, 1988);
N. Sanchez, hep-th/9711068.

\bibitem{cooper}
F. Cooper, A. Khare and U. Sukhatme,
Phys. Rep.  {\bf 251}, 267(1995).

\bibitem{uvarov} A. F. Nikiforov and V. B. Uvarov,
{\it Special Functions of Mathematical Physics},
(Birkh\"{a}user, Boston, 1988).


\end{thebibliography}
\end{document}